# Active disturbance rejection for precise positioning of dual-stage hard disk drives

Mohammad Amin Dashti
Dept. of mechanical engineering
University of Guilan
Rasht, Iran
m.amin.dashti@gmail.com
m.amin.dashti@birjand.ac.ir

Ali Chaibakhsh
Dept. of mechanical engineering
University of Guilan
Rasht, Iran
chaibakhsh@guilan.ac.ir

*Abstract*— **This paper presents an application of adaptive control algorithm in order to reject the external disturbances in dual-stage hard disk drives. For this purpose, a dual PID controller is first designed without the plant exposure to external disturbances. Then, an adaptive control approach based on recursive least squares adaptive (RLS) algorithm was employed to identify and reject disturbances. The performance of the proposed technique was evaluated for hard disk track-seeking through simulation experiments. Results show the feasibility and precise tracking of the designed control system.**

*Index Terms*__ Recursive Least Squares Algorithm, feedforward Control, dual-stage hard disk drives, external disturbance, identification, voice coil motor, micro actuator.

## I. Introduction

Dual-stage actuator has been employed as employed for many years as the main track-reading mechanism in magnetic hard-disk drive for achieving higher storage capacities[1]. The large speed with high displacement and high precision for a unique actuator is not attained due to mass inertia. The usage of dual-stage manipulators in robotics industry is a technic to reach to high precision and fast response. By joining a voice coil motor (VCM) with a large travel range but poor precision to a micro-actuator with higher precision and faster response along with an appropriate control design, the drawbacks of two actuators can be compensated by each other [2]. An adaptive control scheme for disk file track-following single stage servo systems was proposed by Roberto Horowitz and Bo Li [3]. Seeking and settling controllers for dual actuator systems have been studied by Kobayashi and Horowitz [4]. An adaptive accelerometer was applied by Pannu and Horowitz [5, 6]. In this paper, the proposed approach by Bando et al. was adopted to deal with external disturbances[7].

## II. System schematic

The most probable source of disturbances on the hard disk drives are: Vibrations and external shock, track misregistration due to bearing hysteresis and poor velocity estimation, inaccuracies and pattern nonlinearities, resonance effects in mechanical parts, electronic noises, non-repeatable run-outs of the spindle, repeatable run-outs due to thermal and other drifts[8].

In order to improve the track-seeking and track-following capabilities, the bandwidth of the head positioning servo system should be increased to reduce the sensitivity of the closed loop system as it is exposed to disturbances such as spindle motor run-out, windage, and disk flutter and external vibrations[9].This paper focuses on isolating the effects of external shock and vibrations that could be measured using accelerometer or velocity sensors.

For modeling a dual-stage hard disk and defining the location of disturbances applying, many different options are available. As it is shown in Fig. 1, it is considered that external vibrational disturbances are applied before micro actuator[10]. The transfer functions for VCM and micro-actuator models are suggested as below[11, 12]:

$$G_V(s) = 3.548 \times \frac{10^7}{s^2} \times \frac{5.536 \times 10^8}{s^2 + 1280s + 5.536 \times 10^8} \quad (1)$$

$$G_M(s) = 0.366 \times \frac{2.975 \times 10^9}{s^2 + 2450s + 1.7 \times 10^9} \times \frac{s^2 + 4524s + 2.08 \times 10^9}{s^2 + 6032s + 3.64 \times 10^9} \quad (2)$$

## III. Controller design

Design a simple and robust controller for VCM and micro-actuator in order to fast tracking and avoid resonance conditions are the main goals in this stage. In track-seeking mode, the major tracking operation is carried out by VCM for larger distances. In cases that the initial head position is far from to desired track position, the main movement of drive head will be performed by VCM. For track-following mode, the micro-actuator performs the main role. In this case, after a short period of time, VCM will stabilize on the desired track path [13, 14]. In order to prevent probable errors and to grantee the safety of micro-actuator, a saturation block is required to limit controller forces on the micro-actuator.

Table 1. Adjustment and specifications of parallel form PID controllers

| PID controller | P | I | D | N (filter coefficient) | Overshoot (%) | Settling time (sec) | Rise time (sec) |
|---|---|---|---|---|---|---|---|
| VCM PID | 0.0328608 | 0.8955647 | 9.86285e-05 | 3316.4 | 13 | 0.00529 | 0.000321 |
| Micro-actuator PID | 0.0650849 | 4.7032010 | 1.99346e-04 | 1402745 | - | 0.00162 | 2.79e-09 |

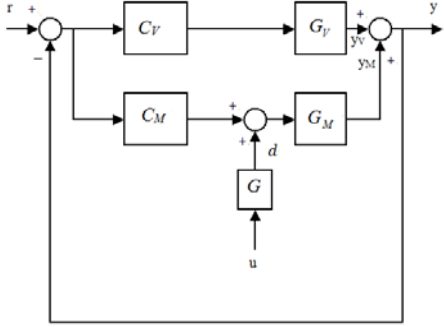

Fig. 1. Block diagram of the dual-stage hard disk drive

To achieve better control system performances, the controllers for VCM and micro-actuator should be designed are two separate stages. First, by considering the VCM and micro-actuator from platform a PID controller is tuned to desired conditions like low overshoot such as 10 percent and low settling time. After designing controller for VCM, the VCM omits from platform and a PID is designed for micro-actuator. So by designing just two PID controllers, the platform will be controlled cheaply and robustly. Fig. 2 shows the step response after tuning controllers in track-following mode control. The tuned quantities of PID controllers for VCM and microactuator are also shown in Table 1.

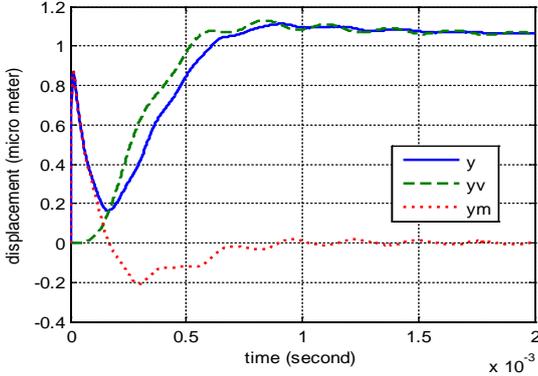

Fig. 2. Track-following mode control at low distance y = 1 μm.

## IV. RLS ALGORITHM

The structure of entering external disturbance is considered as Fig. 1. External disturbance from the base frame to micro-actuator has a dynamic tranftsfer function $G(z^{-1})$. By adaptive RLS Algorithm, this transfer function is identified at the track-following mode. At the track-following mode, step command is adjusted to zero. Upon the disturbance identification, transfer function of disturbance will be feedforward. RLS algorithm can be formulated as:

$$G = \frac{B(z^{-1})}{A(z^{-1})} \qquad (3)$$

$$A(z^{-1}) = 1 + a_1 z^{-1} + a_2 z^{-2} + \ldots + a_n z^{-n} \qquad (4)$$

$$B(z^{-1}) = b_1 z^{-1} + b_2 z^{-2} + \ldots + b_m z^{-m} \qquad (5)$$

By considering $d(k)$ as the transfer function of plant disturbances, we have:

$$d(k) = \frac{B(z^{-1})}{A(z^{-1})} u(k) + e(k) \qquad (6)$$

$$d(k) = -a_1 d(k-1) - a_2 d(k-2) - \ldots - a_n d(k-n) + b_1 u(k-1) + b_2 u(k-2) \ldots + b_m u(k-m) \qquad (7)$$

$$\theta^T = [a_1, a_2, \ldots, a_n, b_1, b_2, \ldots, b_n] \qquad (8)$$

$$\varphi^T(k) = [-d(k-1), \ldots, -d(k-n), u(k-1), \ldots, u(k-m)] \qquad (9)$$

$$\hat{d}(k) = \hat{\varphi}^T(k)\hat{\theta}(k-1) \qquad (10)$$

By defining the error as the difference between real and identified disturbances:

$$e(k) = d(k) - \hat{d}(k) \qquad (11)$$

$$\hat{\theta}(k) = \hat{\theta}(k-1) + \frac{P(k-1)\varphi(k)}{1 + \varphi^T(k)P(k-1)\varphi(k)} e(k) \qquad (12)$$

$$\lambda(k) = 1 - \frac{P(k-1)\varphi(k)^2}{1 + \varphi^T(k)P(k-1)\varphi(k)} \frac{1}{trP(0)} \qquad (13)$$

$$P(k) = \frac{1}{\lambda(k)} \{P(k-1) - \frac{P(k-1)\varphi(k)\varphi^T(k)P(k-1)}{1 + \varphi^T(k)P(k-1)\varphi(k)}\} \qquad (14)$$

Here $\hat{\theta}(k)$ is the identified parameter and $\varphi^T(k)$ is the signals of the input and the output of the identified disturbance transfer function. $d(k)$ is the output signal of real disturbance transfer function and $\hat{d}(k)$ is the output signal of identified disturbance transfer function. $\lambda(k)$ is called forgetting factor to make estimation smoother. When $\lambda = 1$ and fixed, the estimates become smoother till the gain P(k) goes to zero. When $\lambda < 1$ the estimator gain P(k) does not go to zero and the estimates will always fluctuate. Here $\lambda$ can be assumed 1 and fixed for fast calculations, although by the Fixed Trace $\lambda(k)$ can be updated as Eq. (11). The different recursive algorithms are quite similar. They can all be described by the Eq. (14) and [15]:

$$\hat{\theta}(k) = \hat{\theta}(k-1) + P(k)\varphi(k)e(k) \quad (15)$$

It should be noted that for the Eq. (10) we have modified $\hat{\varphi}^T(k)$ instead of $\varphi^T(k-1)$ while Astroom has described in regression model the Eq. (10) as below [15]:

$$\hat{d}(k) = \varphi^T(k-1)\hat{\theta} \quad (16)$$

## V. DESIGN FEEDFORWARD

The parameters of disturbance transfer function are updated by spending just a few numbers of iterations. Initial conditions are important to converge the estimated parameters to the true parameters. The rotational acceleration of the disc drive body is sensed and applied to an adaptive filter that produces a feedforward signal (Fig. 3) designed to offset the effects of the rotational vibration. The adaptive filter adjusts its parameters based on the rotational acceleration signal, the position error signal of the servo system, and a transfer function relating the actual position signal to the feedforward signal [16].

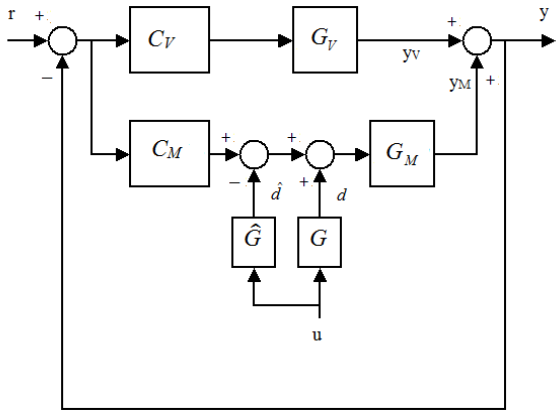

Fig. 3. Adaptive feedforward signal design

## VI. SIMULATION RESULTS

For external disturbance simulation, it is considered as a sine wave with $10^4$ micro meter amplitude and different frequencies like 100 and 200 Hz. The selecting the random disturbance with high magnitude will be rejected also; because both of them excite the system dynamics adequately. Sampling interval is selected 0.2 milli second. The order of $A(z^{-1})$ and $B(z^{-1})$ are selected $n_a = n_b = 2$ to make calculations faster and without complexity. It is said in [15] and some other references $P(1) = 10 - 10000 \, I_{na+nb}$, so for a fast converging error to zero it has selected $P(1) = 10^4 \, I_4$ ( "$I_4$" is an identity matrix with 4×4 order.). Initial $\hat{\theta}(1)$ is not much important, but it enhances the algorithm iteration times a few more such as 20 to 50 iterations. Other assumptions are as below:

$$\hat{\theta}^T(1) = [0,0,0,0]$$

$$G = \frac{z^{-1} + z^{-2}}{1 + z^{-1} + 0.5z^{-2}}$$

The results are presented by this algorithm and compared in in Fig. 4 and Fig. 5 for 100 Hz vibration. Fitness of identified disturbance transfer function by proposed algorithm for 100 Hz vibration is 87.45 %. Clearly it is observed that disturbance effect has decreased to 5 micrometer under high amplitude 100 Hz vibration in Fig. 5. After spending 50 iterations (10 milli second) identified transfer function parameters are converged to fixed parameters with a good precision.

$$\hat{G} = \frac{-0.302z^{-1} + 1.106z^{-2}}{1 + 0.2426z^{-1} + 0.1016z^{-2}}$$

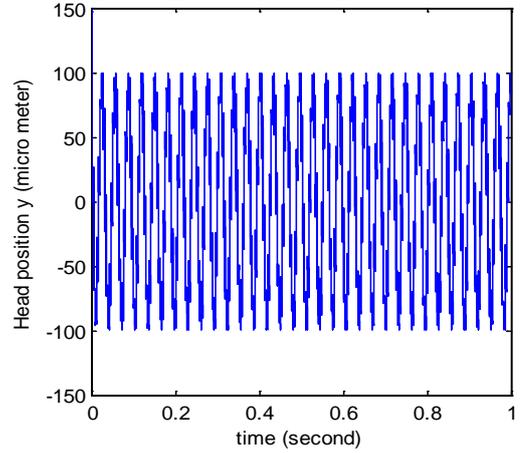

Fig. 4. Head position y under 100 Hz vibration without feedforward signal

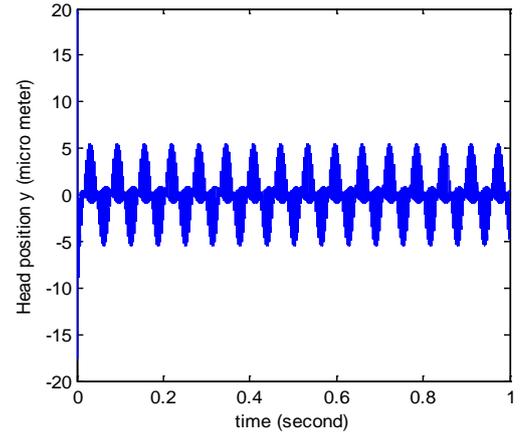

Fig. 5. Head position y under 100 Hz vibration with feedforward signal

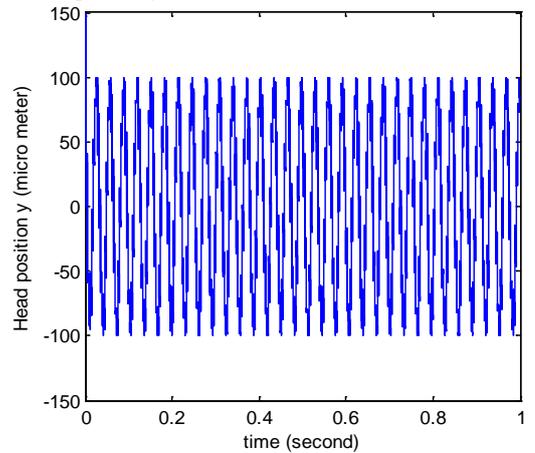

Fig. 6. Head position y under 200 Hz vibration without feedforward signal

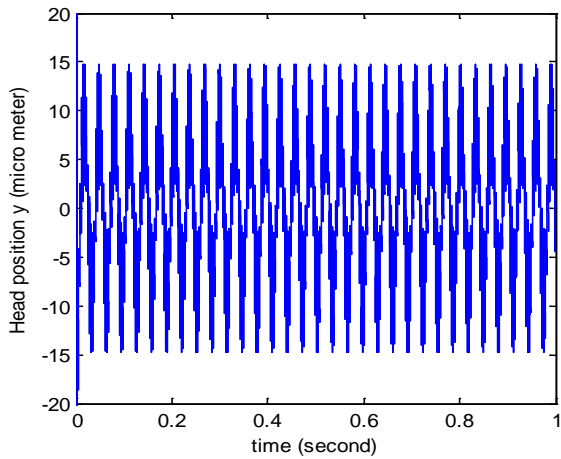

Fig. 7. Head position y under 200 Hz vibration with feedforward signal

Fitness of identified disturbance transfer function by proposed algorithm for 200 Hz vibration is also 74.99 %. The disturbance effect has decreased to 15 micrometer under high amplitude 200 Hz vibration based on Fig. 6 and Fig. 7.

## VII. Conclusion

We extend in this paper external disturbance rejection for a Dual-stage hard disk drive, reducing effects of vibration by rotational acceleration measurement. For identification of dynamic disturbance transfer function, adaptive Recursive Least Squares algorithm is applied. So we reach to a precise and better performance due to implementing the micro-actuator. The track-following becomes more precise than using just a voice coil motor actuator. The variation of dynamics of disturbance input till base frame and head is not also concerning because adaptive Recursive Least Squares identifies disturbance transfer function parameters effectively. The estimated disturbance transfer function is fed forward to reject disturbance effects. The controller design is simple, cheap and also effective. For Dual-stage hard disk servo mechanism, dual PID controllers are designed which are more reliable and have simpler appearance with respect to other methods. The details of Adaptive Recursive Least Squares algorithm implementation is explained clearly.